\documentclass[%
 reprint,
 amsmath,amssymb,
 aps,
floatfix,
]{revtex4-2}
\usepackage{dcolumn}

\usepackage[caption=false]{subfig}
\usepackage{amssymb}
\usepackage{amsmath}
\usepackage{commath}
\usepackage{graphicx,bm}
\usepackage{verbatim}

\begin{document}

\raggedbottom

\title{Narrow and broadband single-photon sources using customised-tapered waveguides}

\author{Harrison R. Greenwood}
 \email{hg35@hw.ac.uk}
 
\author{Mohammed F. Saleh}%
 \email{m.saleh@hw.ac.uk}
\affiliation{Institute of Photonics and Quantum Sciences, Heriot-Watt University, EH14 4AS Edinburgh, United Kingdom}

\date{\today}

\begin{abstract}
In this paper, we present a thorough investigation for a spontaneous parametric four-wave mixing process in third-order nonlinear waveguides with various continuous tapering patterns. It has been previously shown that these devices can quasi-phase-match the four-wave-mixing process and enhance its conversion efficiency by orders of magnitude. By altering the tapering profile curve we found that these devices can enable single-photon sources with either narrow or broadband spectral widths at on-demand frequencies. Using our model, we were also able to identify the waveguide length at which the single-photon spectral purity is maximised. 
\end{abstract}

\maketitle

\section{Introduction}

Quantum schemes using single-photons provide a feasible route to achieve strong advances in various fields such as communication and computation systems. Single-photons can be deterministically generated via quantum dots \cite{Kako2006,Bennett2005,Senellart2017}, colour centres \cite{Beveratos2002,Khramtsov2018}, and atomic ensembles \cite{Chou2004,Matsukevich2006}, or probabilistically via spontaneous parametric down-conversion (SPDC) \cite{Burnham1970,Zhong2010,Kaneda2018} and spontaneous four-wave-mixing (SFWM) processes \cite{Dyer2008,Takesue2007,Christ2009} in nonlinear optical media.

Spontaneous processes have become the standard practice in generating single photons for quantum applications due to the ease of implementation at standard laboratory conditions \cite{Spring2017}. Preparing desirable photon states is typically achieved in second-order media via SPDC, \cite{Fedrizzi2007,Graffitti2018_2} a three-wave-mixing process that converts a parent photon into two photons at a lower frequency. This process is inherently constrained by energy and momentum conservation \cite{fund_saleh}, the latter of which leads to the phase-matching condition. Overcoming the limitations imposed by the phase-matching condition is key for efficient photon production. This is typically achieved by implementing periodically-poled crystals \cite{qpm_review_tuningtolerances} where the sign of the second-order nonlinear coefficient is alternated over discrete domains enabling power to flow efficiently from the pump field into the generated field. This process is possible because second-order nonlinear media lack an inversion centre. In third-order media the preparation of quantum states of light can be achieved through SFWM, a parametric process where two pump photons are combined to create two other output photons. A number of techniques have previously demonstrated the capacity to satisfy the phase-matching condition in third-order media including photonic-crystal fibres \cite{Rarity2005,Garay-Palmett2007}, microresonators \cite{Vernon2017}, birefringent waveguides \cite{Spring2017,Spring2013}, and directional couplers \cite{Dong2004,Francis-Jones2018}. 

Periodically tapered waveguides (PTW) have also recently been proposed as a means to quasi-phase-match (QPM) parametric nonlinear processes in third-order nonlinear materials \cite{qpm_supercontinuum, saleh_1}. These structures are analogous to periodically-poled crystals \cite{qpm_review_tuningtolerances}. In third-order nonlinear media, which possess inversion symmetry, a similar variation of the nonlinear coefficient can be achieved through a periodic modulation of the waveguide cross-section. The fibre core diameter or waveguide width is continuously modulated along the propagation direction \cite{Driscoll2012, Armaroli2012}. We found that by carefully choosing values of the tapering period, \(\Lambda_{\mathrm{T}}\), and modulation amplitude, \(\Delta\sigma\), high conversion efficiency can be achieved.

An exciting prospect for continuously tapered devices is in enabling efficient generation of single-photons at on-demand frequencies in third-order nonlinear materials. Launching an optical pulse into modulated guiding structures results in the production of photon-pairs facilitated by nonlinear mixing processes. This technique overcomes the current limitations of parametric processes in third-order nonlinear materials by enabling the production of widely spectrally-separated photons. It has been previously demonstrated that the PTW-technique results in enhanced efficiency for third-harmonic generation \cite{saleh_1} as well as spontaneous-four-wave-mixing \cite{saleh_2}. Here we have implemented periodic and nonperiodic tapering patterns to tailor the spectral properties of photons spontaneously emitted via four-wave-mixing in silica-core photonic-crystal fibres. We have found that periodic patterns result in efficient narrow-band photon-pairs while nonperiodic structures lend themselves to broad bandwidth generation. We have also thoroughly investigated the effect of the proposed structures on the generated photons spectral purity, and study the effect of fabrication intolerances.

This paper is organised as follows. Sec. \ref{model} outlines the quantum model developed specifically in order to analyse tapered devices for single-photon generation using continuous and pulsed pump sources. In Sec. \ref{simulations} we have applied this model by simulating SFWM processes in photonic-crystal fibres with various tapering profiles. Finally, our conclusions are summarised in Sec. \ref{conclusion}.

\section{Modelling}\label{model}

We adapt a recently developed numerical approach based in the Heisenberg picture \cite{saleh_2} to study parametric SFWM processes in a single-mode third-order nonlinear medium. This approach divides the nonlinear medium into discrete segments of equal thickness. Within a given segment the linear and nonlinear properties are assumed to be constant, provided that the segment thickness is small relative to the tapering period. The spatial evolution of photons within each segment is described by a set of coupled differential equations whose solutions can be written in a matrix form, repeating this process for each segment a transfer matrix for the entire device can be determined. Subsequently the expected number of photons, \(\langle N\rangle\), for a particular combination of the signal and idler frequencies can be calculated. This model is well suited for describing tapered structures, in comparison to the common interaction Hamiltonian model \cite{photon_pairs_1}, as it accounts for spatially dependent refractive indices and nonlinear coefficients. Importantly, this model has added the benefit of tracking the spectral purity of photons as they propagate along the device as well as accounting for the effect of cross- (XPM) and self-phase modulation (SPM). The Raman nonlinearity has been neglected in this analysis since the PTW-technique can allow for photon-pair production far from the Raman gain peak \cite{Armaroli2012}.

\subsection{Electric fields of interacting photons}\label{efield_evolution}

The pump (\(p\)) source is assumed to be a strong undepleted optical pulse where its electric field can be described as a superposition of multiple monochromatic waves,

\begin{equation}\label{pumpelectricfield}
    \mathcal{E}_p(x,y,z,t) = \sum_v E_v(x,y,z,t) + \mathrm{c.c.},
\end{equation}
with \(x\) and \(y\) the transverse coordinates, \(z\) the longitudinal direction, \(t\) the time coordinate, and \(\mathrm{c.c.}\) the complex conjugate. The complex variable \(E_v=\frac{1}{2}A_vF_v(x,y)e^{-j(w_vt-k_vz)}\), where \(A_v\), \(F_v\), \(\omega_v\), and \(k_v\) are the amplitude, transverse profile, angular frequency, and propagation constant of a monochromatic wave \(v\) respectively, and \(k_v=n_v\omega_v/c\) with \(n_v\) the refractive index at a frequency \(\omega_v\) and \(c\) the speed of light in a vacuum. 

The propagation of each individual monochromatic pump wave through a uniform nonlinear waveguide facilitates the flow of power from the pump field into the signal and idler (\(s\), \(i\)) fields through a SFWM process. The electric field operator of generated signal/idler photons can be defined as,

\begin{equation}\label{outputelectricfield}
    \hat{\mathcal{E}}(x,y,z,t) = \hat{\mathcal{E}}^+(x,y,z,t) + \hat{\mathcal{E}}^-(x,y,z,t),
\end{equation}
where \(\hat{\mathcal{E}}^+\) and \(\hat{\mathcal{E}}^-\) are the positive and negative frequency components of the field, and \(\hat{\mathcal{E}}^-=(\hat{\mathcal{E}}^+)^{\dagger}\). The positive frequency component \(\hat{\mathcal{E}^+}\) can be described as a superposition of frequency-dependent mode operators,

\begin{equation}\label{frequencysuperposition}
    \hat{\mathcal{E}}^+(x,y,z,t) = \sum_s \sqrt{\frac{\hbar \omega_s}{2\epsilon_0cTn_sS_s}}F_s(x,y)\hat{a}(z,\omega_s)e^{-j\omega_st},
\end{equation}
where \(\hbar\) is the reduced Planck's constant, \(\epsilon_0\) is the dielectric permittivity, \(T\) is a short time period, \(S_s =\int\int|F_s|^2dxdy\) is the beam area, \(\hat{a}\) is the annihilation operator, and \(\omega_s\), \(n_s\), \(F_s(x,y)\) are the angular frequency, refractive index, and transverse profile of a particular mode, \(s\). 

The spatial evolution of the signal/idler electric field operators are governed by the commutation relation \cite{parametric_amplifier_analysis,shen_quantum_statistics},

\begin{equation}\label{spatial_evolution}
    -j\hbar\frac{\partial\hat{\mathcal{E}}}{\partial z} = [\hat{\mathcal{E}},\hat{G}],
\end{equation}
where \(\hat{G}\) is the momentum operator defined over a time window \(T\), as,

\begin{equation}\label{big_G}
    \hat{G}(z)=\int\int\int_0^T \hat{g}(x,y,z,t)dtdxdy, 
\end{equation}
with the momentum-flux operator \(\hat{g}=\hat{\mathcal{D}}^+\hat{\mathcal{E}}^- + \mathrm{H.c.}\), \(\hat{\mathcal{D}}\) the electric-displacement field operator given by \(\hat{\mathcal{D}} = \epsilon_0 n^2 \hat{\mathcal{E}} + \hat{\mathcal{P}}_{N\!L}\), \(\hat{\mathcal{P}}_{N\!L}\) the nonlinear polarisation operator, and \(\mathrm{H.c.}\) is the Hermitian conjugate. Equation \ref{spatial_evolution} provides a relatively straightforward method of determining the evolution of the field operator which in turn gives the evolution of the annihilation operator \(\hat{a}\). 

\subsection{Signal and idler coupled mode equations}

For each monochromatic source any nonlinear mixing process is restricted by the simple relation, \(2\omega_v = \omega_s + \omega_i\). This is a result of energy conservation and it restricts the photon frequencies that can be generated. Considering this condition alongside contributions from the linear propagation, cross-phase modulation (XPM), and four-wave-mixing (FWM) process, we obtain coupled mode equations that describe the evolution of both the creation and annihilation operators of the signal and idler photons \cite{saleh_2},

\begin{equation}\label{annihilation_coupled}
    \frac{\partial\hat{b}_s}{\partial z} = j\gamma_{s,i} e^{j\Delta\phi}\hat{b}^\dagger_i(z),
\end{equation}
\begin{equation}\label{creation_coupled}
    \frac{\partial\hat{b}^\dagger_i}{\partial z} = -j\gamma_{s,i} e^{j\Delta\phi}\hat{b}_s (z),
\end{equation}
where \(\hat{b}^\dagger\) and \(\hat{b}\) are the phase transformed creation and annihilation operators respectively, \(\Delta\phi = \int_0^z\Delta\kappa(z')dz'\) is the accumulated phase-mismatch, and \(\Delta\kappa = 2\kappa_p - \kappa_s - \kappa_i\) is the phase mismatch. The nonlinear coefficient \(\gamma_{s,i}\) and the propagation constants are given by,

\begin{equation}\label{gammasi}
    \gamma_{s,i} = \frac{3\chi^{(3)}A_p^2}{4c}\sqrt{\frac{\omega_s\omega_i}{n_sn_iS_sS_i}}\int\int F_p^2F_s^*F_i^*dxdy.
\end{equation}
\begin{equation}\label{kappap}
    \kappa_p = k_p\left(1 + \frac{3\chi^{(3)}A_p^2}{8n_p^2S_p}\int\int|F_p|^4dxdy\right),
\end{equation}
\begin{equation}\label{kappas}
    \kappa_u = k_u\left(1 + \frac{3\chi^{(3)}A_p^2}{2n_u^2S_u}\int\int|F_p|^2|F_u|^2dxdy\right),
\end{equation}
where \(\chi^{(3)}\) is the third-order susceptibility, \(u = s,i\) and \(k_q= n_q\omega_q/c\) with \(q = s,i,p\). The parameter \(\kappa_p\) accounts for self-phase modulation in the pump beam, while \(\kappa_u\) accounts for cross-phase modulation between the pump and the signal/idler photons. 

Equations \ref{annihilation_coupled} and \ref{creation_coupled} can be solved by dividing the waveguide into discrete elements with constant cross-sections. The operators evolution can then be calculated step by step through the use of a transfer matrix,

\begin{equation}
    {\begin{bmatrix}\hat{b}_s \\ \hat{b}_i^{\dagger}\end{bmatrix}}_{z=z_m+\Delta z} = \mathcal{T}_m{\begin{bmatrix}\hat{b}_s \\ \hat{b}_i^{\dagger}\end{bmatrix}}_{z=z_m},
\end{equation}
with,

\begin{equation}
    \mathcal{T}_m = {\begin{bmatrix}1 & f_{s,i}(w_p) \\ f_{s,i}^*(w_p) & 1\end{bmatrix}}_{z=z_m},
\end{equation}
where \(f_{s,i}(w_p) = j\gamma_{s,i}\Delta ze^{j\Delta\phi}\) and \(\Delta z\) is the element thickness. Each of these matrices \(\mathcal{T}_m\) describes a single element of the waveguide. The matrix for the entire structure can be described for a certain combination of two modes \(s\) and \(i\) as \(\textbf{T}_{s,i}=\mathcal{T}_M\mathcal{T}_{M-1}...\mathcal{T}_2\mathcal{T}_1\), with \(M\) total number of elements. For a specific mode \(s\) the expected number of photons \(\langle N\rangle\) at the output of the device is given by \(\langle\phi|\hat{N}(L,\omega_s)|\phi\rangle = \langle\phi|\hat{b}_s^{\dagger}(L)\hat{b}_s(L)|\phi\rangle = |\textbf{T}_{s,i}(1,2)|^2\), with \(|\phi\rangle=|0\rangle_s|0\rangle_i\) the initial quantum state and \(L\) the waveguide length.

To describe the evolution of two given coupled modes \(s\) and \(i\) in the signal spectrum under the influence of a pulsed pump we account for every possible contribution from pairs of monochromatic waves \(\omega_{p_1}\) and \(\omega_{p_2}\) within the pump spectrum. This is reflected in the transfer matrix structure as,

\begin{equation}
    \mathcal{T}_m = \begin{bmatrix}1 & \sum\limits_{\omega_{p_1}}f_{s,i}(\omega_{p_1},\omega_{p_2}) \\ \sum\limits_{\omega_{p_1}}f_{s,i}^*(\omega_{p_1},\omega_{p_2}) & 1\end{bmatrix}_{z=z_m}.
\end{equation}
The summation has been reduced from a double summation to one over \(\omega_{p_1}\) by applying the energy conservation relation \(\omega_{p_2}=\omega_s+\omega_i-\omega_{p_1}\).

\section{Simulations}\label{simulations}

In this section the aforementioned model is implemented to investigate photon-pair generation via spontaneous-four-wave-mixing (SFWM) in periodically and non-periodically tapered fibres. In particular, the influence of different tapering patterns, amplitudes, and periods on the efficiency of the SFWM mixing processes and the spectral properties of generated photons have been thoroughly studied. The fibres are designed to operate in the normal-dispersion regime where it is difficult to satisfy the phase-matching condition using uniform single-mode waveguides. This also has the potential to suppress undesirable nonlinear phenomena that can be excited under strong pumping. It is important that the propagation within the tapered structures remains adiabatic. For this reason small modulations of the fibre diameter are coupled with relatively large tapering periods, in comparison to the operating wavelength.

Simulations were performed in silica solid-core photonic-crystal fibres (PCFs) with diameters that vary longitudinally \cite{mod_instability}. These fibres are currently fabricated with tapering periods of a few tens of centimetres and modulation amplitudes of \(10\%\). The cladding of these structures comprises a stack of hollow capillary tubes, with a cross-sectional pitch, \(\sigma\), and a hole-diameter, \(d=0.5\sigma\), as depicted in the inset of Fig. \ref{fig:gaus_figures}(a). We have assumed single-mode operation since this cladding arrangement prevents the formation of higher-order modes that usually experience very high propagation losses. Additionally, the phase-matching condition would prevent the generation of single-photons in undesirable modes. The effective refractive indices are calculated using the Sellmeier equation of silica \cite{fund_saleh} and a set of empirical equations \cite{empirical_pcf}. We have considered PCFs with pitches that vary according to chirped-sinusoidal and Gaussian patterns. The period of the simulated fibres has been discretised into 200 steps to increase the accuracy of the results. All fibres are modelled with an average pitch \(\sigma_\mathrm{av} = 1\,\mathrm{\mu m}\). Finally, we define our fibre profiles according to how the fibre pitch evolves with the longitudinal coordinate, \(z\), since it is approximately proportional to the fibre diameter.

\subsection{PCFs with Gaussian tapering patterns}

\begin{figure}
\includegraphics[width=1\linewidth]{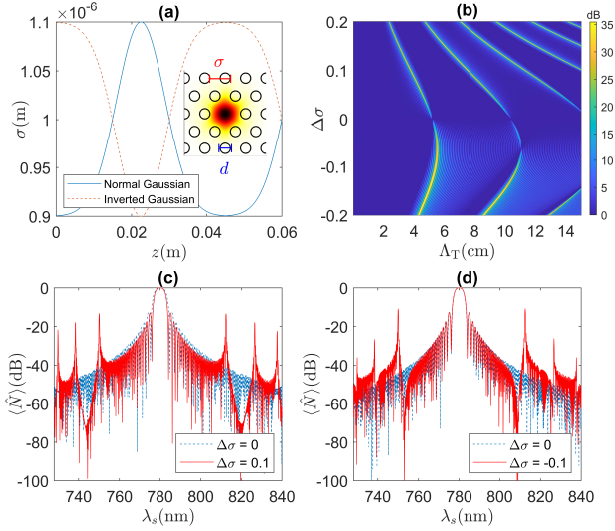}
\caption{\label{fig:gaus_figures} Colour online. (a) Tapering profile for normal (solid blue) and inverted (dashed orange) Gaussian patterns, including a schematic view of the PCF cross-section displaying the mode profile, modulation amplitude \(\sigma\), and hole diameter \(d\), with \(\sigma=1\,\mu \mathrm{m}\), \(\lambda_p=780\,\mathrm{nm}\), and an effective index \(n_{e\!f\!f}=1.42\). (b) Dependence of the expected number of photons \(\langle N\rangle\) on the tapering period \(\Lambda_\mathrm{T}\) and modulation amplitude \(\Delta\sigma\) at \(\lambda_s = 750 \,\mathrm{nm}\) and \(M = 50\) for a Gaussian profile. (c,d) Spectral dependence of \(\langle N\rangle\) for normal Gaussian fibres (\(\Lambda_\mathrm{T} = 3.78 \,\mathrm{cm}\), \(\Delta\sigma = 0.1\), solid red) and inverted Gaussian fibres (\(\Lambda_\mathrm{T} = 5.41 \,\mathrm{cm}\), \(\Delta\sigma = -0.1\), solid red) with \(M = 50\) periods across the signal spectrum. Uniform fibres (dashed blue) have been added for comparison.}
\end{figure}

To construct a periodic Gaussian function, individual Gaussian functions are truncated and ``stitched'' together. The truncation procedure is performed to ensure a smooth transition between neighbouring curves and avoid scattering losses. Unlike the common sinusoidal patterns, flipping a Gaussian profile results in two asymmetric patterns (normal and inverted) as shown in Fig. \ref{fig:gaus_figures}(a). The figure inset displays the internal PCF structure, indicating the fibre pitch \(\sigma\) and hole diameter \(d\), alongside the pump beams transverse mode profile. For a Gaussian fibre the pitch varies as \(\sigma(z) = 2\Delta\sigma e^{-\frac{1}{2}[(z-\Lambda_\mathrm{T}/2)/w]^2}\), where \(w = \Lambda_\mathrm{T}/7\) is the characteristic Gaussian width and \(\Delta\sigma\) is the modulation amplitude. 

The dependence of the expected number of photons, \(\langle N\rangle\), on the tapering period, \(\Lambda_\mathrm{T}\), and modulation amplitude, \(\Delta\sigma\), for a \(1\,\mathrm{W}\) continuous pump wave at \(780\,\mathrm{nm}\) is shown in Fig. \ref{fig:gaus_figures}(b). Certain combinations of \(\Lambda_\mathrm{T}\) and \(\Delta\sigma\) result in large enhancement of the conversion efficiency by satisfying the quasi-phase-matching condition. These combinations manifest as bright trajectories that depict the \(n^{th}\) order tapering periods from left to right. The values of \(\langle N\rangle\) have been normalised to the \(\Delta\sigma = 0\) case (uniform fibres). Similar to periodically-poled structures, as the tapering-order increases the efficiency of conversion decreases.

In contrast to the sinusoidal case studied in Ref. \cite{saleh_2}, Fig. \ref{fig:gaus_figures}(b) shows that the expected number of photons for normal and inverted Gaussian profiles have an asymmetric dependence on the modulation amplitude, \(\Delta\sigma\). The inverted Gaussian curve produces an expected number of photons with a higher efficiency over a relatively broader range in comparison to the normal Gaussian curve. The cause of this enhanced efficiency can be found by analysing the Fourier spectra of the phase-mismatch term, \(e^{j\Delta\phi}\), and nonlinear coefficient, \(\gamma_{s,i}\) \cite{saleh_1}. We observe a better overlap between the two spectra because the bright trajectories in the inverted case are broader than the normal case, as displayed in Fig. \ref{fig:gaus_figures}(b). This allows the nonlinear coefficient to more effectively counteract the growth of the phase-mismatch.

\begin{figure}
\includegraphics[width=1\linewidth]{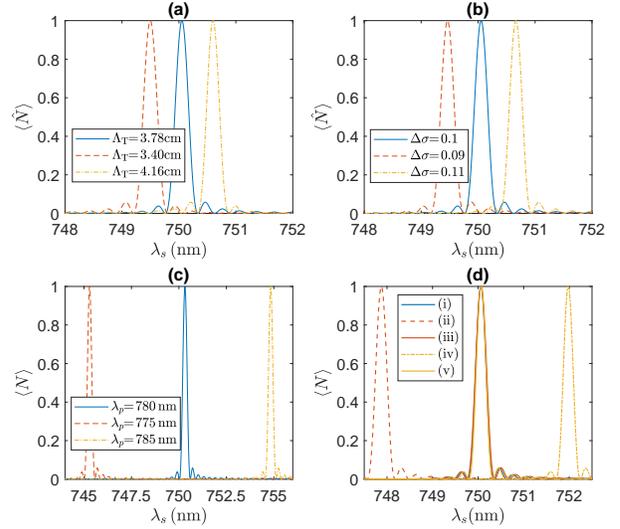}
\caption{\label{fig:enhancement_figures} Colour online. Spectral dependence of \(\langle N\rangle\) around \(\lambda_s\) on tapering period \(\Lambda_\mathrm{T}\), modulation amplitude \(\Delta\sigma\), and pump wavelength \(\lambda_p\) (a) Normal Gaussian fibres pumped at \(780\mathrm{nm}\), \(\Delta\sigma = 0.1\), and \(\Lambda_\mathrm{T}=3.78\mathrm{cm}\pm10\%\). (b) \(\lambda_p=780\!\mathrm{nm}\), \(\Lambda_\mathrm{T}=3.78\mathrm{cm}\), and \(\Delta\sigma=0.1\pm10\%\). (c) \(\Lambda_\mathrm{T}=3.78\mathrm{cm}\), \(\Delta\sigma = 0.1\), \(\lambda_p=780\mathrm{nm}\pm5\mathrm{nm}\). (d) (i): \(\lambda_p=780\mathrm{nm}\), \(\Lambda_\mathrm{T}=3.78\mathrm{cm}\), \(\Delta\sigma = 0.1\). (ii): \(\lambda_p=780\mathrm{nm}\), \(\Lambda_\mathrm{T}=3.40\mathrm{cm}\), \(\Delta\sigma = 0.09\). (iii): \(\lambda_p=782.3\mathrm{nm}\), \(\Lambda_\mathrm{T}=3.40\mathrm{cm}\), \(\Delta\sigma = 0.09\). (iv): \(\lambda_p=780\mathrm{nm}\), \(\Lambda_\mathrm{T}=4.16\mathrm{cm}\), \(\Delta\sigma = 0.11\). (v): \(\lambda_p=778\mathrm{nm}\), \(\Lambda_\mathrm{T}=4.16\mathrm{cm}\), \(\Delta\sigma = 0.11\).}
\end{figure}

The 1D spectral dependence of \(\langle N\rangle\) on the signal wavelength for a normal and an inverted Gaussian are displayed in Fig. \ref{fig:gaus_figures}(c) and \ref{fig:gaus_figures}(d), respectively. In both cases we compare the results with the spectrum obtained from a uniform fibre. In the uniform case \(\langle N\rangle\) rapidly decays as the signal frequency moves away from the pump wavelength at \(780\,\mathrm{nm}\). In comparison the modulated Gaussian fibre spectra reproduce the classical nonlinear modulation instability effect, which validates the applied model. In both the normal and inverted cases we observe multiple high gain peaks generated away from the central wavelength, indicating that high conversion efficiency can be achieved far from the pump frequency using the PTW-technique. The spectral band around \(750\mathrm{nm}\) in the inverted Gaussian fibre is found to generate approximately \(2.79\times10^9\) photons per second after \(50\) periods, offering photon generation rates close to those found in SPDC schemes \cite{Schneeloch2019}, which shows the potential for our structures. This number increases to \(1.12\times10^{10}\) photons for \(200\) periods.

Figure \ref{fig:enhancement_figures} shows the spectral band for normal Gaussian fibres with a variety of different physical parameters to study the effect of fabrication tolerance. Figures \ref{fig:enhancement_figures}(a,b) show that the central wavelength is shifted to the right and left by approximately \(\pm0.5\mathrm{nm}\) by changing the tapering period or the modulation amplitude by \(\pm10\%\). The previously modelled Gaussian fibre, with \(\Lambda_\mathrm{T}=3.78\mathrm{cm}\) and \(\Delta\sigma=0.1\), acts as a reference point. Figure \ref{fig:enhancement_figures}(c) depicts the effect of tuning the pump wavelength by \(\pm5\mathrm{nm}\). While the observed shift in Figs. \ref{fig:enhancement_figures}(a,b) is \(<1\mathrm{nm}\), by tuning the pump wavelength we observe a shift of \(\approx5\mathrm{nm}\). Figure \ref{fig:enhancement_figures}(d) depicts how the pump wavelength can be tuned to accommodate the fabrication tolerance. Curve (i) is the ideal case. In comparison, curves (ii) and (iv) are for normal Gaussian fibres with both tapering period and modulation amplitude adjusted by \(-10\%\) each, and \(+10\%\) each, respectively. In order to compensate for the changes in the physical parameters of each fibre, the pump wavelength has been tuned so as to generate photons at the desired wavelength. This results in curves (iii) and (v), pumped at \(782.3\mathrm{nm}\) and \(778\mathrm{nm}\) respectively. 

\begin{figure}
\includegraphics[width=1\linewidth]{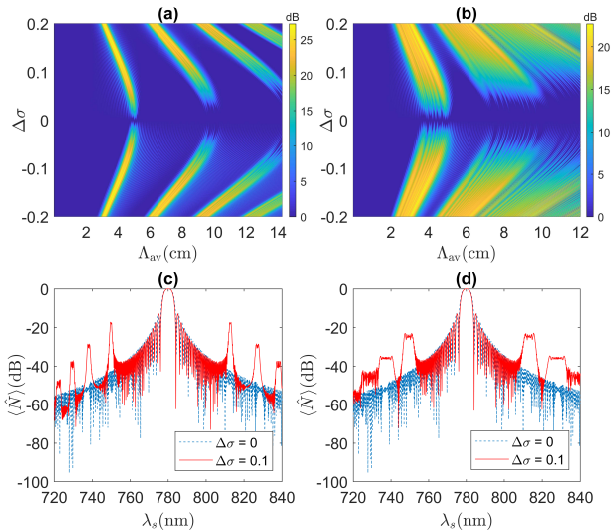}
\caption{\label{fig:chirp_figures} Colour online. (a,b) Dependence of the expected number of photons \(\langle N\rangle\) on the average tapering period \(\Lambda_\mathrm{av}\) and modulation amplitude \(\Delta\sigma\) at \(\lambda_s = 750 \,\mathrm{nm}\) in a chirped-sinusoid fibre after \(M = 50\) with \(C = -6.22\,\mathrm{m^{-2}}\) and \(-25.85\,\mathrm{m^{-2}}\), respectively. (c,d) Spectral dependence of \(\langle N\rangle\) in chirped-sinusoid fibres (solid red) with \(\Lambda_\mathrm{av} = 4.50 \,\mathrm{cm}\) and \(\Delta\sigma = 0.1\) on the signal wavelength \(\lambda_\mathrm{s}\) after \(50\) periods with \(C = -6.22\,\mathrm{m^{-2}}\) and \(C = -25.85\,\mathrm{m^{-2}}\), respectively. Uniform fibres have been added for comparison (dashed blue).}
\end{figure}
 
\subsection{PCFs with chirped-sinusoidal tapering patterns} \label{chirpedsection}

In this subsection we investigate PCFs with a sinusoidal profile, however, they are linearly chirped according to, \(\sigma(z) = \sigma_\mathrm{av}[1 - \Delta\sigma \cos{(f_iz + \frac{1}{2}Cz^2)]}\), where \(f_i = 2\pi/\Lambda_i\) is the initial spatial frequency, \(\Lambda_i\) is the initial tapering period, \(C = (f_f - f_i)/L\) is the chirp parameter, \(f_f = 2\pi/\Lambda_f\) is the final spatial frequency, \(\Lambda_f\) is the final tapering period, and \(L\) is the fibre length. The dependence of the expected number of photons on the average tapering period and modulation amplitude for two different PCFs with weak and strong chirp parameters after \(50\) periods is portrayed in Fig. \ref{fig:chirp_figures}(a) and \ref{fig:chirp_figures}(b) respectively. The bright trajectories are still present, however, they become broader due to satisfying the QPM condition over multiple spectral frequencies. Increasing the chirp size results in increased spectral width at the expense of reducing the expected number of photons. 

Figures \ref{fig:chirp_figures}(c) and \ref{fig:chirp_figures}(d) display the spectral dependence of \(\langle N\rangle\) on the signal spectrum, \(\lambda_\mathrm{s}\), for chirped-sinusoid fibres with the weak and strong chirp parameters respectively. Comparing these spectra to the Gaussian profiles in Figs. \ref{fig:gaus_figures}(c,d), it is clear that the chirp has the effect of broadening the gain peaks but with a decreased conversion efficiency. In other words a wider range of frequencies are able to satisfy the phase-matching condition in a chirped-sinusoid fibre but fewer photons are generated at a given frequency. 

\subsection{Linear variation of the modulation amplitude} \label{varampsection}

\begin{figure}
\includegraphics[width=1\linewidth]{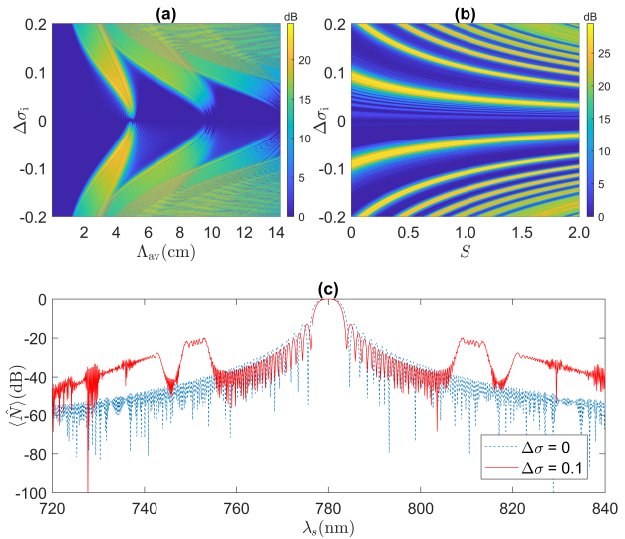}
\caption{\label{fig:varamp_figures} Colour online. (a) Dependence of \(\langle N\rangle\) on the modulation amplitude \(\Delta\sigma\) and the average tapering period \(\Lambda_\mathrm{av}\). (b) Dependence of \(\langle N\rangle\) on the initial modulation amplitude \(\Delta\sigma_\mathrm{i}\) and scaling parameter \(S\). (c) Full spectrum dependence of \(\langle N\rangle\) on the signal wavelength \(\lambda_s\) for an amplitude varying chirped-sinusoid fibre (solid red) with \(\Lambda_\mathrm{av} = 4.5\,\mathrm{cm}\), \(\Delta\sigma = 0.1\), \(S = 1\), \(C = -6.22\,\mathrm{m^{-2}}\) with 50 periods. A uniform fibre (dashed blue) has been added for comparison.}
\end{figure}

Here, we investigated the dependence of the output photon properties on the modulation amplitude. We have introduced a scaling parameter, \(S\), such that an \(S\) value of \(0\) corresponds to chirped-sinusoid fibres with a constant modulation amplitude along the length of the fibre, while \(S = 2\) corresponds to a fibre with a modulation amplitude that increases linearly from \(\Delta\sigma_\mathrm{i}\) to \(3\Delta\sigma_\mathrm{i}\) along the propagation direction. The fibres investigated here are scaled with \(S = 1\) corresponding to a final modulation amplitude \(2\Delta\sigma_\mathrm{i}\). 

Unlike fibres with constant modulation amplitude the separate trajectories that correspond to the right combinations of \(\Delta\sigma\) and \(\Lambda_\mathrm{av}\) merge together, as shown in Fig. \ref{fig:varamp_figures}(a). The broadening becomes more pronounced at greater values of \(\Lambda_\mathrm{av}\) while the conversion efficiency drops as in the case of the chirped-sinusoid. This is again the result of photons generated in fewer quantities over a broad range of frequencies. Figure \ref{fig:varamp_figures}(b) displays the dependence of \(\langle N\rangle\) on the initial modulation amplitude and scaling parameter. The bright solutions take on a symmetric set of curves, similar to Fig. \ref{fig:varamp_figures}(a), that become broader at the extremes of \(\Delta\sigma_\mathrm{i}\) and \(S\). Figure \ref{fig:varamp_figures}(c) shows the signal spectrum of amplitude-varied chirped-sinusoid fibres when pumped continuously at \(780\,\mathrm{nm}\) with \(1\,\mathrm{W}\) input power. The spectrum is composed of very broad gain peaks with decaying amplitudes.

\begin{figure}
\includegraphics[width=1\linewidth]{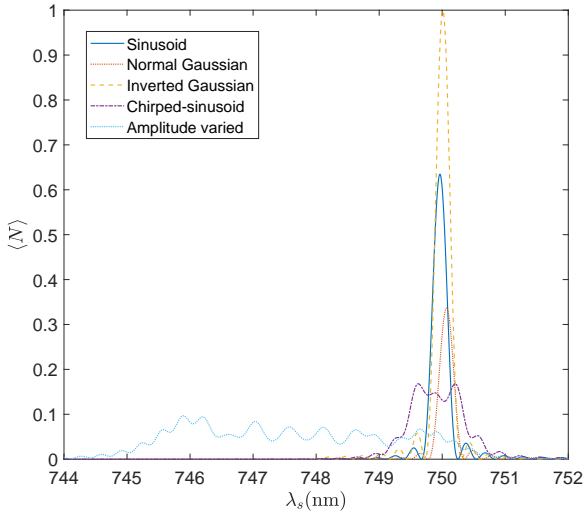}
\caption{\label{fig:comp_enhancement} Colour online. Spectral dependence of \(\langle N\rangle\) on \(\lambda_\mathrm{s}\) at \(M = 50\) for sinusoid (solid blue), normal Gaussian (dotted orange), inverted Gaussian (dashed yellow), chirped-sinusoid (dashed-dotted purple), and amplitude varied (dotted blue) fibre profiles. Each data set is normalised to the inverted Gaussian maximum.}
\end{figure}

\subsection{Discussion}

The normalised expected number of single photons generated in PCFs, used in the above sections, as well as for a PCF with sinusoidal tapering pattern are displayed together for comparison as portrayed in Fig. \ref{fig:comp_enhancement}. The highest conversion efficiency is achieved using the inverted Gaussian pattern in comparison to the sinusoidal and normal Gaussian cases. We found that this is due to better overlap between the Fourier components of the nonlinear coefficient and phase-mismatch term in this case. The careful alignment of these two spectra is essential for efficient QPM, each component of the nonlinear coefficient is utilised in counteracting the growth of a corresponding term in the phase-mismatch.

Additionally, it is clear from Fig. \ref{fig:comp_enhancement} that the application of a chirped pattern results in a large spectral broadening at the expense of lower conversion efficiency. This can be understood as many signal/idler frequencies being phase-matched over different small segments of the structure. This can also be seen in the wide range of \(\Lambda_{\mathrm{T}}\) and \(\Delta\sigma\) combinations that produce high conversion efficiency at a particular wavelength for a chirped sinusoidal tapering pattern, as portrayed in Figs. \ref{fig:chirp_figures}(a) and \ref{fig:chirp_figures}(b).

Spectral purity is another important feature that characterises single photons. It is a measure of the dis-correlation between the single-photons of the generated pairs. In the presented model, the transfer matrix element \(\textbf{T}_{s,i}(1,2)\) is used to construct the joint-spectral amplitude function \(\textbf{J}(\omega_s,\omega_i)\). This function is a product of the phase-matching function and the pump's spectral envelope. To calculate the spectral purity, \(J\) is then factorised via the singular-value decomposition into three matrices \(U\), \(V\), and \(W\). The key parameter in determining the spectral purity is \(V\), a diagonal matrix describing the mixing of modes in the unitary matrices \(U\) and \(W\). Normalising \(V\) such that \(\sum_i\rho^2_i = 1\), with \(\rho_i\) the diagonal elements of \(V\), the spectral purity can be calculated, \(\sum_i\rho^4_i\) \cite{Mosley2008a,Graffitti2018_2}.

\begin{figure}[t]
\includegraphics[width=1\linewidth]{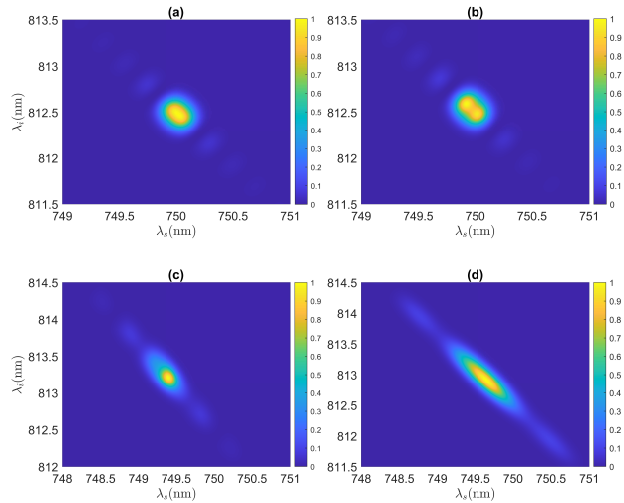}
\caption{\label{fig:purity_figures} Colour online. Joint spectral intensity function for various fibres at the point of maximum purity. (a,b) Dependence of \(\langle N\rangle\) on signal and idler wavelengths at \(M = 75\) for normal (\(\Lambda_\mathrm{T} = 3.78\,\mathrm{cm}\), \(\Delta\sigma = 0.1\)), and inverted (\(\Lambda_\mathrm{T} = 5.41\,\mathrm{cm}\), \(\Delta\sigma = -0.1\)) Gaussian fibres. (c) Dependence of \(\langle N\rangle\) on signal and idler wavelengths at \(M = 39\) for a chirped-sinusoid fibre with \(\Lambda_\mathrm{av}=4.5\,\mathrm{cm}\), \(\Delta\sigma=0.1\), and \(C = -3.11\,\mathrm{m^{-2}}\). (d) Dependence of \(\langle N\rangle\) on signal and idler wavelengths at \(M = 22\) for a amplitude varied chirped-sinusoid fibre with \(\Lambda_\mathrm{av}=4.5\,\mathrm{cm}\), \(\Delta\sigma_\mathrm{i}=0.1\) \(C = -3.11\,\mathrm{m^{-2}}\) and \(S = 1\).}
\end{figure}

The joint-spectral intensity function of the aforementioned microstructured fibres are displayed in Fig. \ref{fig:purity_figures} for a Gaussian pump pulse with an input energy \(1\,\mathrm{nJ}\) and full-width-half-maximum \(4\,\mathrm{ps}\). 
The JSI for normal and inverted Gaussian fibres are shown in Figs. \ref{fig:purity_figures}(a,b) for the optimal number of periods at which the purity reaches a maximal value. The maximum purity of photons generated in Gaussian fibres is found to be very close to \(P = 0.83\) and \(P = 0.81\) in the normal and inverted Gaussian fibres, respectively. Although the inverted Gaussian results in a higher conversion efficiency, the output photon is marginally less pure in comparison to the normal profile. The JSI for the chirped-sinusoid tapering pattern with \(C = -3.11\,\mathrm{m^{-2}}\) is displayed in Fig. \ref{fig:purity_figures}(c), with \(M=39\) being the period at which a maximum purity of \(P = 0.67\) occurs. We found that increasing the number of periods beyond this value would modify the JSI by introducing multiple peaks. This deteriorates the purity of output photons (for instance \(P = 0.29\) at \(M = 100\)), diminishing the usefulness of chirped tapering patterns in generating highly pure photons in relatively long fibres. However, these sources can be exploited in applications that require broadband entangled photons such as quantum optical coherence tomography \cite{Nasr2003}. Figure \ref{fig:purity_figures}(d) shows the JSI for a linearly-varied modulated fibre. In this case, the JSI is similar to that of the chirped fibre. The linear variation of \(\Delta\sigma\) results in large bandwidth that comes at the cost of low spectral purity. The maximum purity achieved in this fibre is \(0.42\) at \(M = 22\) periods.

\begin{figure}
\includegraphics[width=0.95\linewidth]{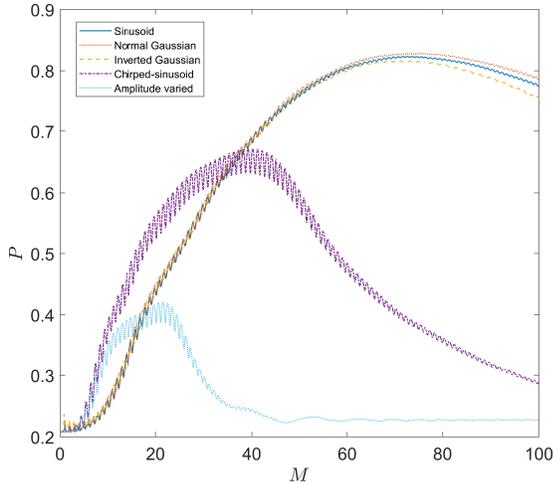}
\caption{\label{fig:comp_purity} Colour online. Purity spatial evolution over the first \(100\) periods for sinusoid (solid blue), normal Gaussian (dotted orange), inverted Gaussian (dashed yellow), chirped-sinusoid (dashed-dotted purple), and amplitude varied (dotted blue) fibre profiles.}
\end{figure}

The spatial dependence of the spectral purity on the number of periods \(M\) for different tapering patterns is depicted in Fig. \ref{fig:comp_purity}. For the periodic structures (sinusoid, Gaussian profiles) a maximum purity \(P = 0.83\) occurs in the normal Gaussian profile at \(M = 74\) while the sinusoidal and the inverted Gaussian achieve slightly lower purities of \(P = 0.82\) at \(M = 73\) and \(0.81\) at \(M = 72\), respectively. Based on these results there is an optimum structure length where the purity is maximised. To improve the spectral purity beyond this maximum, either an idealised phase-matching process via alteration of \(\Lambda_\mathrm{T}\) and \(\Delta\sigma\) or optical filters are needed. Each periodic fibre profile displays a similar level of degradation with the fibre length after the maximum is reached. This is due to the accumulation of a small phase-mismatch, a result of the tolerance in the tapering period, which in-turn results in degradation of the spectral purity at longer distances. The chirped pattern behaves similarly to the periodic patterns but with a lower maximum purity \(P = 0.67\) and a rapid rate of degradation as the device length increases past \(40\) periods. Likewise, the fibre with a linear variation in the modulation amplitude behaves much like the chirped-sinusoid, however the degradation of the purity is further increased and falls rapidly after only \(22\) periods, having achieved a maximum value of \(0.42\).

\section{Conclusion}\label{conclusion}

The presented results suggest that tapered fibres can offer a new potential scheme for tailored photon-pair generation via spontaneous four-wave mixing processes in third-order nonlinear materials. The tapering profile can be chosen to enable higher conversion efficiency over either narrow- or broad-band on-demand frequencies, while tuning of the pump wavelength can be exploited in order to compensate for fabrication tolerances. We found that the spectrum of the output photon is broadened in fibres with nonperiodic patterns at the cost of having lower conversion efficiency. In addition, using our model, we were able to identify the optimum number of periods needed to increase the output photon spectral purity. 
We envisage implementing tailored tapering patterns could lead to higher efficiency and spectral purities, comparable to those achieved in periodically-poled SPDC structures. Also, the tapering pattern can be used to produce single-photons with same frequencies but different polarisations, as commonly done in SPDC systems. Moreover, this technique can be applied in planar waveguides \cite{qpm_supercontinuum}, this would help in advancing the rapidly-evolving integrated-quantum-photonics research field \cite{integrated_pqt_review}. Finally, we anticipate that complex patterns can yield further desirable spectral properties, which in-turn will produce exciting avenues of new research.

\clearpage

\bibliographystyle{apsrev4-1}

\end{document}